\documentclass[journal=jacsat,manuscript=article]{achemso}

\usepackage[version=3]{mhchem} 
\usepackage[dvipsnames]{xcolor}

\author{Angelina Frank}
\email{angelina.frank@u.nus.edu}

\affiliation[Centre for Quantum Technologies]
{Centre for Quantum Technologies, National University of Singapore, Singapore}

\author{Justin Zhou}
\email{justinzhou@u.nus.edu}
\affiliation[Department of Physics]
{Department of Physics, National University of Singapore, Singapore}

\author{James A. Grieve}
\affiliation[IIT]
{Quantum Research Centre, Technology Innovation Institute, Abu Dhabi}

\alsoaffiliation[Centre for Quantum Technologies]
{Centre for Quantum Technologies, National University of Singapore, Singapore}

\author{Ivan Verzhbitskiy}
\affiliation[Department of Physics]
{Department of Physics, National University of Singapore, Singapore}

\author{Jos\'{e} Viana-Gomes}
\affiliation[Departamento de Física, Centro de Física]
{Departamento de Física, Centro de Física, Campus de Gualtar, Portugal}

\author{Leyi Loh}
\affiliation[Department of Physics]
{Department of Physics, National University of Singapore, Singapore}

\author{Michael Schmid}
\affiliation[4th Physics]
{4th Physics Institute and Research Center SCoPE, University of Stuttgart, Germany}

\author{Kenji Watanabe}
\affiliation[Tsukuba]
{Research Centre for Functional Materials, National Institute for Materials Science, Japan}

\author{Takashi Taniguchi}
\affiliation[Tsukuba2]
{International Centre for Materials Nanoarchitectonics, National Institute for Materials Science, Japan}

\author{Goki Eda}
\email{g.eda@nus.edu.sg}
\affiliation[Department of Physics]
{Department of Physics, National University of Singapore, Singapore}
\alsoaffiliation[Chemistry]
{Department of Chemistry, National University of Singapore, Singapore}
\alsoaffiliation[CA2DM]
{Centre for Advanced 2D Materials, National University of Singapore, Singapore}

\author{Alexander Ling}
\affiliation[Centre for Quantum Technologies]
{Centre for Quantum Technologies, National University of Singapore, Singapore}
\alsoaffiliation[Department of Physics]
{Department of Physics, National University of Singapore, Singapore}

\title[An \textsf{achemso} demo]
  {Mode-center Placement of Monolayer WS$_{2}$ in a Photonic Polymer Waveguide}

\begin{document}


\begin{abstract}
  Effective integration of 2D materials such as monolayer transition metal dichalcogenides (TMDs) into photonic waveguides and integrated circuits is being intensely pursued due to the materials’ strong exciton-based optical response. Here, we present a platform where a WS$_{2}$ -hBN 2D heterostructure is directly integrated into the photonic mode-center of a novel polymer ridge waveguide. FDTD simulations and collection of photoluminescence from the guided mode indicate that this system exhibits significantly improved waveguide-emitter coupling over a previous elastomer platform. This is facilitated by the platform’s enhanced refractive-index contrast and a new method for mode-center integration of the coupled TMD. The integration is based on a simple dry-transfer process that is applicable to other 2D materials, and the platform’s elastomeric nature is a natural fit to explore strain-tunable hybrid-photonic devices. The demonstrated ability of coupling photoluminescence to a polymer waveguide opens up new possibilities for hybrid-photonic systems in a variety of contexts.
\end{abstract}

\section{TOC Graphic}

\begin{figure}
 \centering
    \includegraphics[width=8cm]{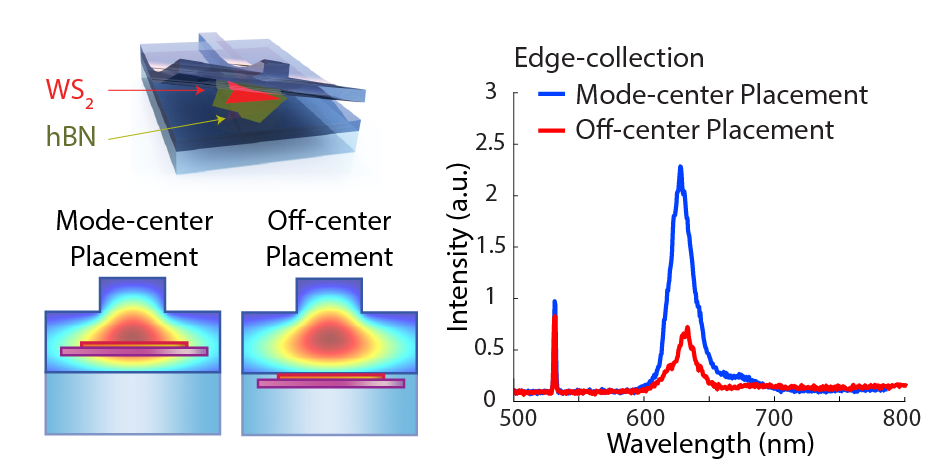}
\end{figure}

\section{Introduction}
Monolayer transition metal dichalcogenides (TMDs) are a material class of great interest due to their exciton-based optical properties in both linear and non-linear regimes\cite{malard_observation_2013,yin_edge_2014,mak_photonics_2016,sun_optical_2016}. They have also been shown capable of hosting single-photon emitters\cite{srivastava_optically_2015,koperski_single_2015,tran_room-temperature_2017}. Besides the tunability of their bandgap in response to electric fields\cite{chu_electrically_2015,wang_electronics_2012}, TMDs’ exciton dynamics can be further modified through stacking into van der Waals heterostructures\cite{rivera_interlayer_2018}, chemical doping\cite{mouri_tunable_2013}, and application of mechanical strain\cite{castellanos-gomez_local_2013}. These properties are of particular interest in the context of Photonic Integrated Circuits (PICs) where the goal is to couple a photon-routing structure to elements with strong and tunable optical response\cite{elshaari_hybrid_2020}. Such circuits find applications in optical communication\cite{sibson_chip-based_2017}, computation\cite{flamini_photonic_2018} and sensing \cite{chamanzar_hybrid_2013,kohler_biophotonic_2021,porcel_invited_2019}. Integration of TMDs into photonic waveguides can be readily achieved by direct transfer, taking advantage of the  material’s van der Waals bonding nature\cite{tonndorf_-chip_2017,joshi_transition_2020}. To date, most hybrid systems with an optically coupled 2D material realize placement away from the photonic mode maximum  (off-center placement) or at a dielectric interface, e.g. close to an exposed fiber core\cite{ngo_scalable_2020,zuo_optical_2020}, at the end-face of an optical fiber\cite{vogl_room_2017}, or on top of a waveguide in integrated photonics\cite{peyskens_integration_2019,he_low-loss_2021,iff_strain-tunable_2019,errando-herranz_resonance_2021,kim_hybrid_2020}.\par
Previously, we have demonstrated integration of a TMD into an elastomeric ridge waveguide\cite{auksztol_elastomeric_2019}. However, the optical coupling efficiency in this system was limited by the material’s off-center placement in between high- and low-refractive index layers and a relatively low index contrast (\emph{n$_{1}$-n$_{2}$ \raisebox{-0.9ex}{\~{}}} 0.001). Addressing this limited coupling efficiency is desirable, since elastomeric waveguides are an exciting platform for hybrid photonic systems due to elasticity, as well as a simple, mild and fast fabrication cycle that is cost effective and favours integration with additional micro- and nanostructures. Potential applications include a.o. rapid-protoyping, foundational material studies and lab-on-a-chip systems\cite{perez-calixto_fabrication_2017, shabahang_single-mode_2021,kee_design_2009,missinne_stretchable_2014,grieve_mechanically_2017}. An additional advantage is the environmental shielding that encapsulation provides for emitters.\par
Here, we present a novel elastomer ridge waveguide together with a new technique to integrate a 2D heterostructure (WS$_{2}$-hBN) in the mode-center of the waveguide (Figure 1 a). The device displays markedly enhanced emitter-mode coupling compared to the previous elastomer platform. The enhancement relies on two factors. Firstly, the newly developed system employs two (commercially available) polymer formulations that increase the index contrast by two orders of magnitude (\emph{n$_{1}$-n$_{2}$ \raisebox{-0.9ex}{\~{}}} 0.1). As a result, the platform can be miniaturized, reducing the effective mode area by a factor of \raisebox{-0.9ex}{\~{}}20 from 166.4 	$\mu$m$^{2}$ for the previous platform down to 8.5 	$\mu$m$^{2}$. This concentration of the electromagnetic field facilitates interaction between material and guided mode, while simultaneously increasing the waveguide’s collection efficiency. Secondly, a new fabrication sequence allows for placement of a 2D material in the mode center. This contrasts with the previous method where a 2D material could only be placed off-center, i.e. between the high- and low-index polymer. Therefore, emission from integrated 2D materials which arises from in-plane dipole sources\cite{schuller_orientation_2013} couples more effectively to the guided mode in the new device. Numerical simulations (FDTD, Lumerical) indicate that this miniaturization and optimized placement lead to an overall improvement of the mode overlap compared to the previous system, and that mode-center placement causes improvement by a factor of 20 compared to off-center placement in the new platform (Figure 1 b and c). This position serves as a benchmark, as it would be the most strongly coupling configuration realizable without the introduced fabrication method. We provide an experimental comparison between mode-center and off-center placement and confirm the improved interaction between mode and 2D material via collection of top-down-excited (free-space-excited) photoluminescence from the guided mode (edge-collection). This collection evidences significant enhancement compared to the previous platform where no signal was observable via edge-collection.

\begin{figure}
 \centering
    \includegraphics[width=17cm]{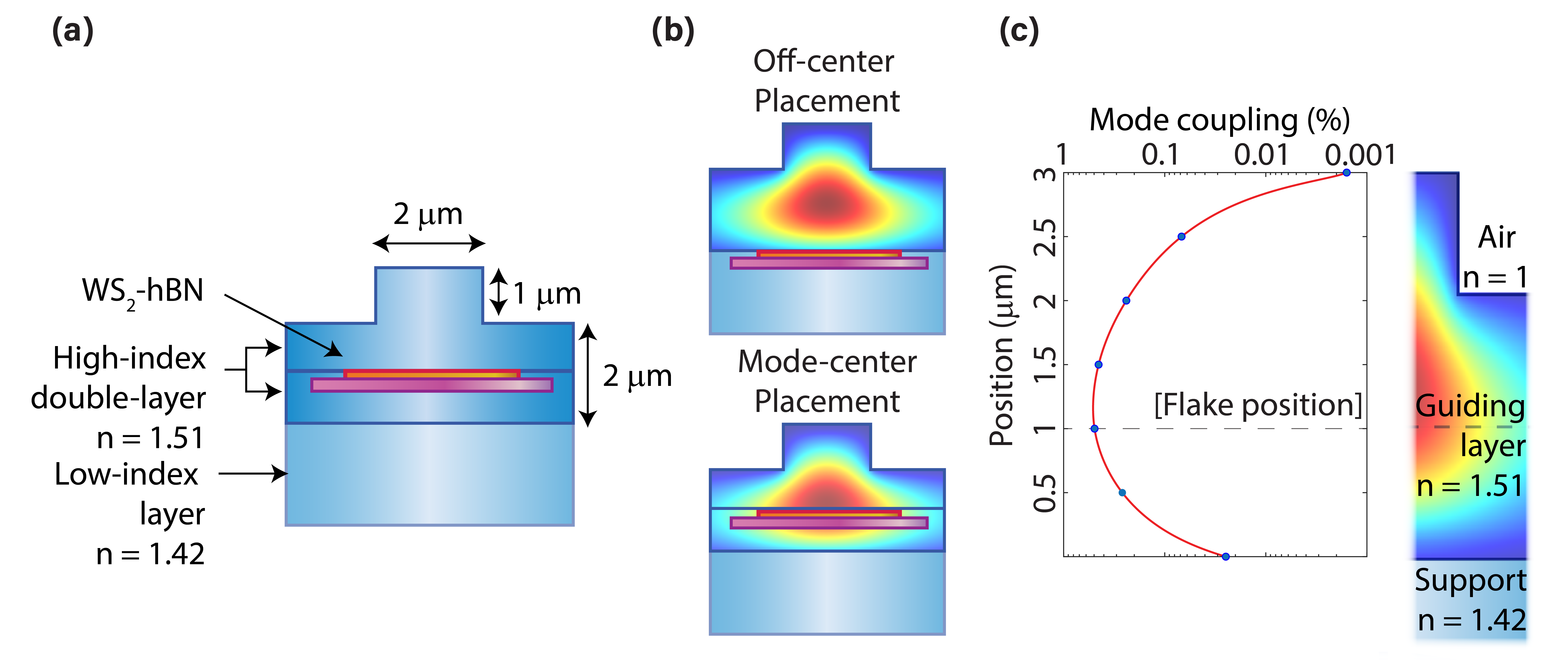}
  \caption{(a) Schematic depiction of the waveguide with the integrated WS$_{2}$-hBN heterostructure (thickness \raisebox{-0.9ex}{\~{}}20 nm) (b) Illustration of off- and mode-center placement (emitter not drawn to scale) together with the mode profile. (c) Simulation of a dipole source’s emission at 620 nm coupling to the waveguide’s fundamental mode for different vertical positioning. A maximum mode coupling of 0.49 \% in one propagation direction is predicted. The red line is intended as a guide to the eye.}
\end{figure}

\section{Experimental}
\begin{figure}
  \includegraphics[width=15cm]{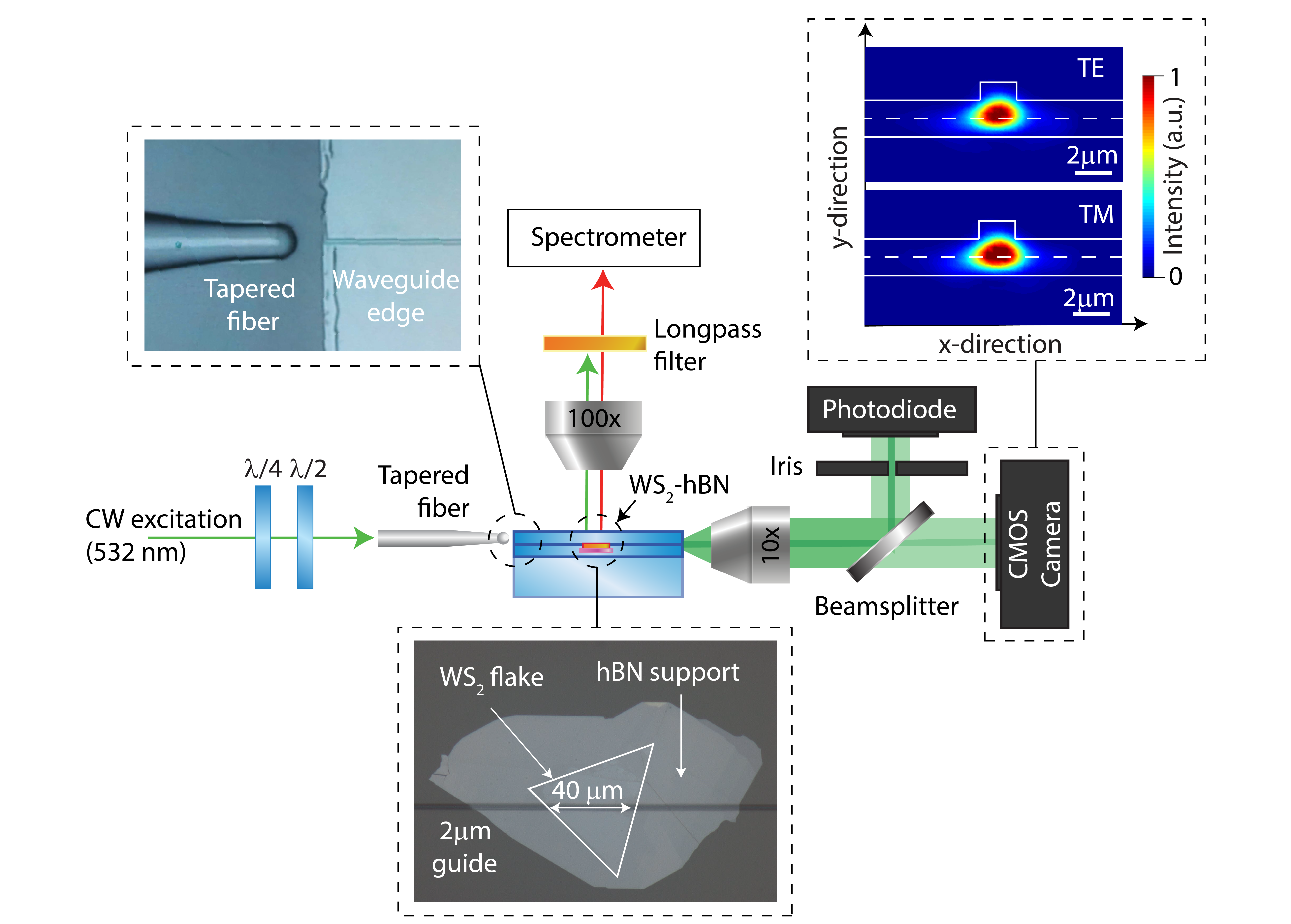}
  \caption{Schematic of the experimental setup configured for edge-excitation with a 532 nm pump laser and collection of photoluminescence from free space. Top left: Microscope image of a tapered fiber aligned to the waveguide for edge-coupling. Top right: Recorded images of the waveguide’s output profile for orthogonal pump polarizations. White lines schematically outline the waveguide. Bottom: Microscope image of the WS$_{2}$-hBN heterostructure within the waveguide. WS$_{2}$ is contained within the white lines.}
\end{figure}
The developed polymer platform is a ridge waveguide based on two polymers (\emph{n$_{1}$ \raisebox{-0.9ex}{\~{}}} 1.42 and \emph{n$_{2}$\raisebox{-0.9ex}{\~{}}} 1.51 at 532 nm). Integration into the mode-center is achieved via dry-transfer of a WS$_{2}$-hBN heterostructure onto a waveguide precursor and subsequent plasma-bonding of two pre-cured polymer layers that make up the lower and the upper half of the high-refractive-index layer (see Methods for details). Each half of this guiding layer is about 1 	$\mu$m thick and the waveguide ridge has a width of 2 	$\mu$m with a height of 1 	$\mu$m (Figure 1a). The structure is capable of single-mode guiding orthogonal polarizations equally (Figure 2, top left). The refractive index difference between guiding polymer layer and the supporting polymer matrix lies between 0.085 to 0.1 and is stable over a range of wavelengths and temperatures as measured by a critical angle refractometry\cite{gissibl_refractive_2017,schmid_optical_2019} (see Supporting Information). \par

A schematic of the experimental setup configured for an edge-coupled pump (edge-excitation) is shown in Figure 2. The polymer substrate is approximately 1.3 cm in length and the WS$_{2}$-hBN heterostructure is placed close to the center of the device. Here, hBN with a thickness of ~20 nm serves as structural support to prevent wrinkling. The WS$_{2}$ monolayer covers a length of 40 	$\mu$m along the waveguide. Fiber, waveguide and collection optics are each mounted onto three- and five-axis translation stages (ULTRAlign, Newport) for alignment. A lensed single-mode fiber (SM-630-HP, conical taper: 1.6-2 	$\mu$m) is edge-coupled on one side and either connected to a spectrometer (NTegra system by NT-MTD) or to a 532 nm pump. A 100x microscope objective (Olympus, 0.6 NA) is focussed onto WS$_{2}$ from the top and connected to either a spectrometer or a 532 nm light source. We linearize the polarization of the edge-coupled fiber using a polarizer, half-wave plate (HWP) and quarter-wave plate (QWP), with the latter pre-compensating for birefringence in the fiber. The edge of the waveguide chip is imaged onto a CMOS camera using a 10x objective lens to facilitate and monitor fiber alignment. An additional non-polarizing beam splitter in this optical collection path allows for simultaneous monitoring of both the transmitted output profile and the transmitted power via a power meter. An iris ensures that only the guided mode’s intensity evolution is recorded. All experiments were performed at room-temperature.

\section{Results}
To characterize optical material-waveguide coupling in the device, we employed FDTD simulations and different excitation-collection schemes as outlined below.
Initially, we estimated how much of the modal power flow (i.e. the Poynting vector along propagation direction) overlaps with the material for different vertical WS$_{2}$ placements inside the waveguide. Modelling WS$_{2}$ as a 1 nm thick and 3 $\mu$m wide area perpendicular to the direction of mode propagation, we calculated a fractional power flow of 0.055\% through the TMD for mode-center placement in contrast to 6.03x10$^{-5}$\% for surface placement and 0.0021\% for placement below the photonic mode i.e. at the interface between high- and low-index polymer which we refer to as off-center placement below. This corresponds to an enhancement of the overlap by almost three orders of magnitude compared to surface placement and one order of magnitude compared to off-center placement.\par

We first confirmed emitter-waveguide coupling by exciting the material from the top and collecting its photoluminescence (PL) from the guided mode via edge-collection (Figure 3). The detection of excitonic PL here is an important illustration of the device’s improved material-guide coupling since the previous system did not show a signal in this configuration\cite{auksztol_elastomeric_2019}. In these experiments, we moved the excitation spot along the waveguide-region covered by WS$_{2}$ and recorded the edge-collected PL for two different devices that realize either mode-center placement or off-center placement (with a material offset from the mode center by 1 $\mu$m).
\begin{figure}
 \centering
    \includegraphics[width=17cm]{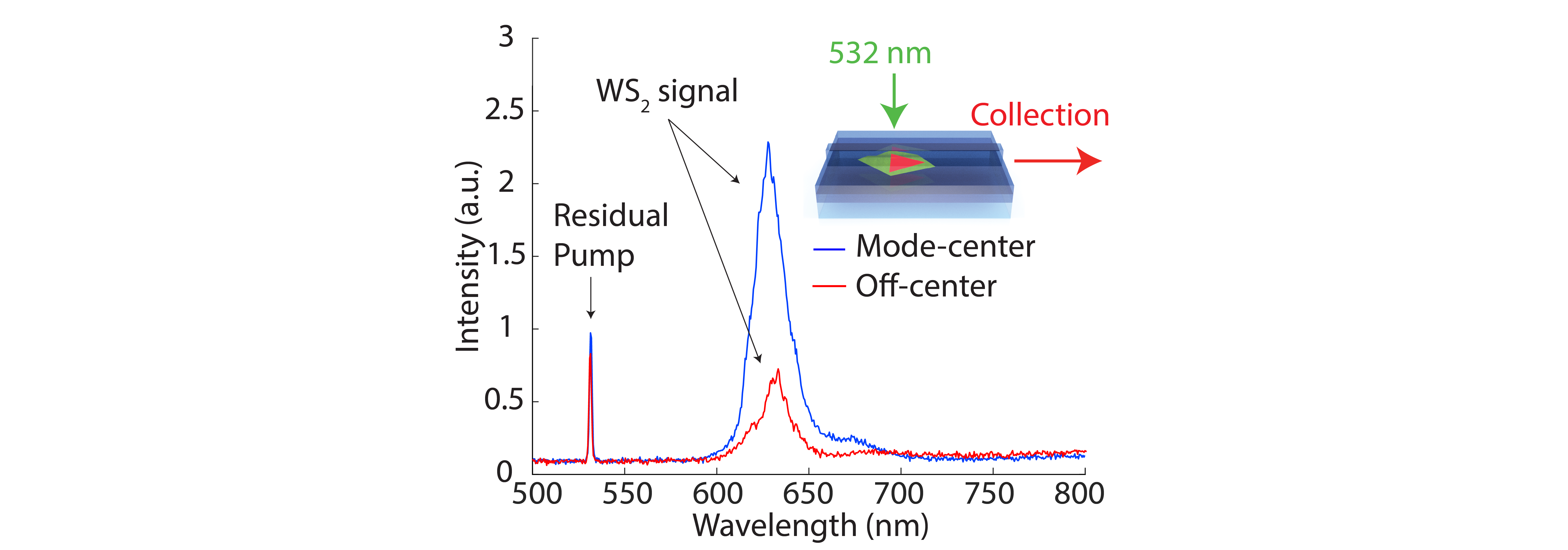}
  \caption{Edge-collected WS$_{2}$ photoluminescence from the guided mode in the developed waveguide system. (a) The stronger WS$_{2}$ photoluminescence coupled to the waveguide via mode-center placement (blue) indicates its superior mode-coupling compared to off-center placement (red) (integration time = 30 s, 350 $\mu$W pump).}
\end{figure}
We note that edge-collection of PL from the waveguide mode is possible for both the mode-center and the off-center device. The difference in coupling predicted to arise from the optimized mode-center placement is indeed observable as a higher average signal. In this configuration, however, a rigorous quantitative evaluation of the relative coupling efficiency is not possible due to setup limitations, and will be subject of future studies.
\par To investigate the relative performance of the mode-center and off-center device more closely, we established a second experimental configuration where excitation and collection points are interchanged such that we recorded the PL emitted into free space upon excitation by an edge-coupled pump (Figure 4 and 5). Using an excitation power of ~40 	$\mu$W ensures that we stay within the linear regime of WS$_{2}$ exciton emission. In this range, the collected PL intensity will be proportional to the excitation rate and serve as a measure of how well the waveguide mode couples to WS$_{2}$. We compared the coupling for mode-center placement with that for off-center placement. To account for local variations in the material's quantum yield, we normalized the PL against the top-down excited and -collected PL at every site. Further, we collected the emission every 2.1 	$\mu$m along the waveguide where it is interfaced with the flake (Figure 4 a). As anticipated, the mode-center device shows a markedly stronger response compared to the off-center device under the same excitation conditions (Figure 4). Moreover, we observed nearly exponential decay of the signal along the channel, indicating extinction of the waveguide mode. We measured an overall signal ratio of 11:1 for mode-center compared to off-center placement. This is evidence for the anticipated facilitation of coupling through mode-center placement, with simulations predicting a \raisebox{-0.9ex}{\~{}}20-times enhancement.  The difference between simulation and experiment may be due to the dielectric properties of the 20 nm thick hBN support which were not included in the model.

\begin{figure}
 \centering
    \includegraphics[width=17cm]{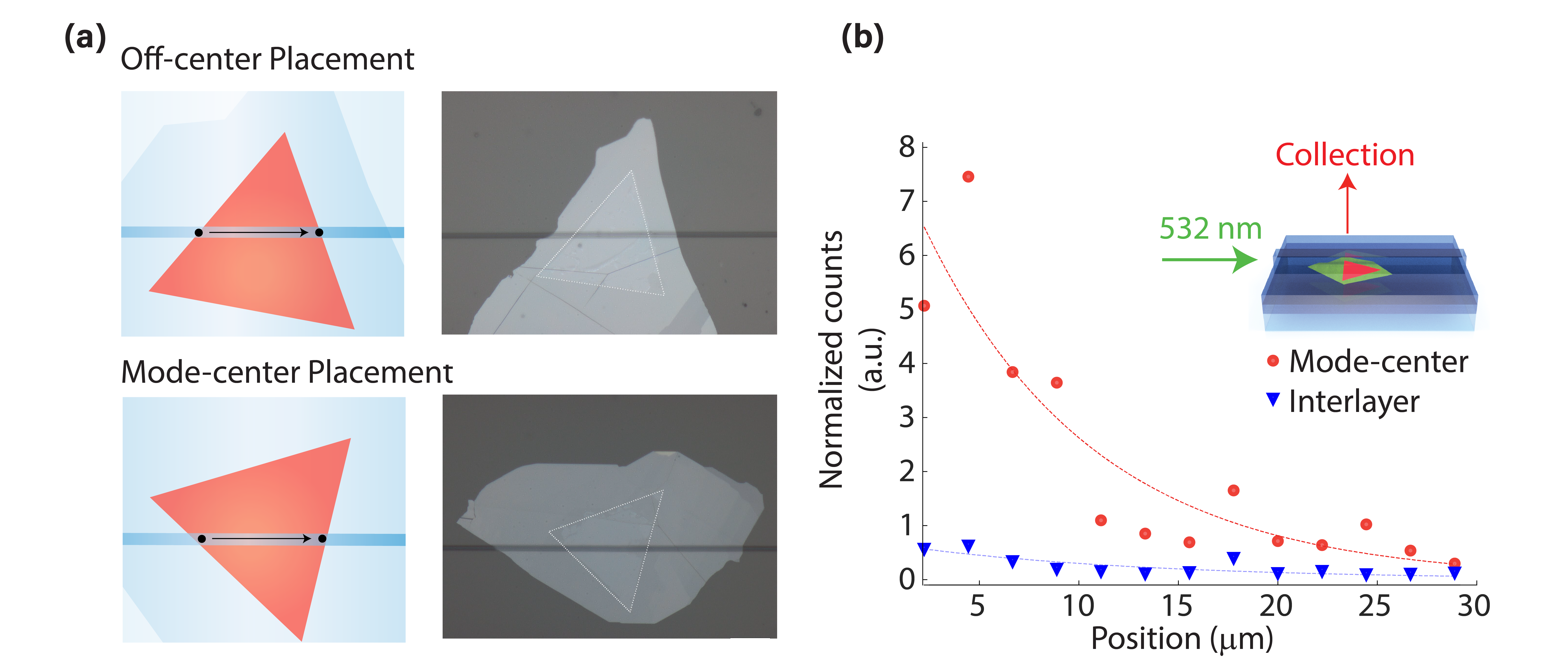}
  \caption{Experimental comparison of mode coupling for WS2-hBN integrated in the mode-center and off-center. (a) The photoluminescence of WS$_{2}$ excited by the photonic mode is collected from free-space at different positions along the waveguide (black dots). (b) Plot of the free-space collected photoluminescence at different positions normalized for local variations in the quantum yield. A least-square exponential fit reveals that the extinction coefficients are similar (0.08 and 0.12) and integration of the counts shows a distinctively higher excitation ratio of 11:1 for mode-center compared to off-center placement.}
\end{figure}

Lastly, to confirm that WS$_{2}$-hBN maintains its structure and that the waveguide is, in fact, polarization-preserving, we recorded the PL in the same scheme but under variation of the pump polarization and at a fixed collection site (Figure 5). This yielded an expected polarization dependence of the PL signal, with an intensity that is minimal for out-of-plane-polarized pump light. The polarization dependence is due to the strictly in-plane orientation of excitons\cite{schuller_orientation_2013}. The extinction ratio, that is, the ratio of the PL intensity for in-plane and out-of-plane polarized excitation, was found to be \raisebox{-0.9ex}{\~{}}65, indicative of this waveguide’s polarization-preserving characteristics. It also serves as a confirmation of the material’s preserved structure throughout fabrication and transfer. A slight asymmetry in the polarization plot originates from experimental imperfections.\par
Weaker coupling of dipole radiation to the waveguide compared to this free-space-collected emission is expected for a dipole oriented in-plane with a Poynting vector that has an angular distribution largely consisting of out-of-plane components. The radiation pattern shows little overlap with the waveguide mode, evoking a weaker signal compared to that collected as free-space radiation. 

\begin{figure}
 \centering
    \includegraphics[width=17cm]{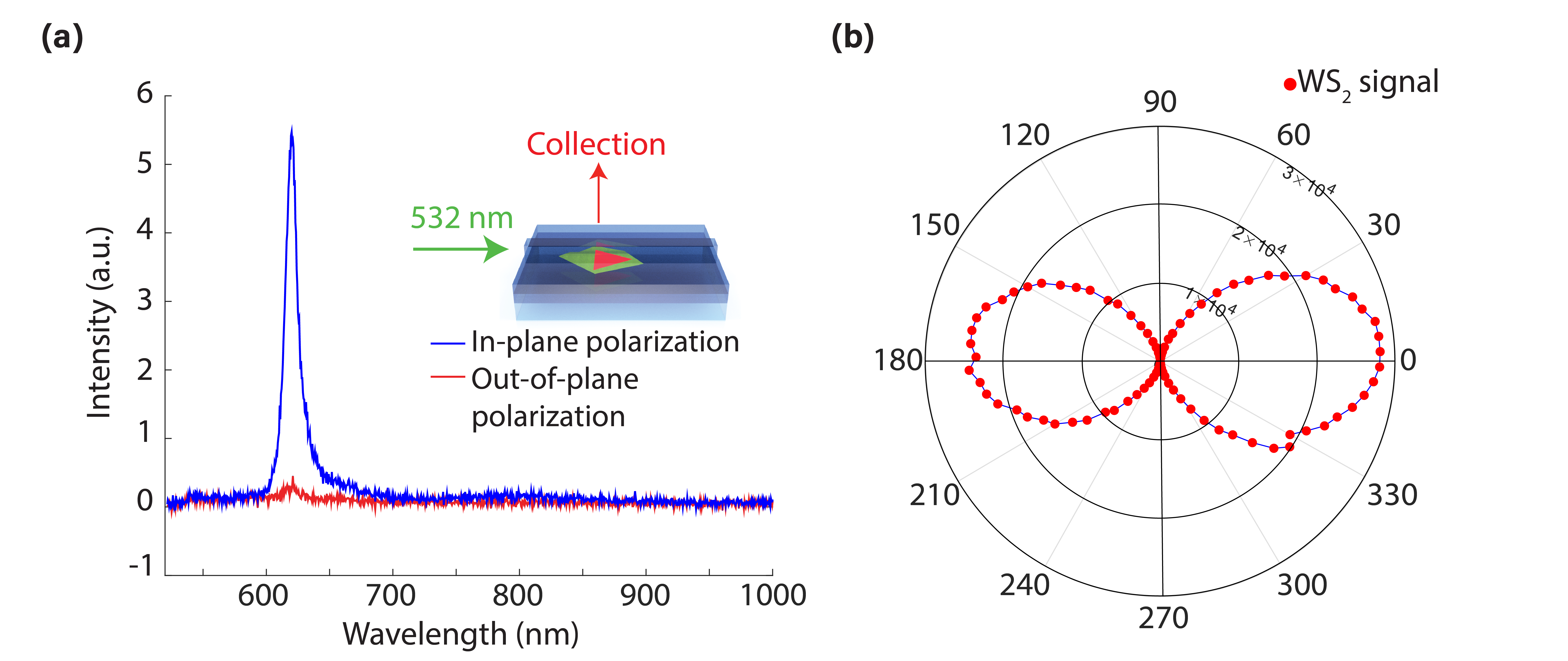}
  \caption{WS$_{2}$ photoluminescence excited by an edge-coupled pump and measured via free-space collection. Out-of-plane polarized light shows minimal interaction with WS$_{2}$. (a) In-plane and out-of-plane polarized pump light couples to the waveguide and evokes different photoluminescence signals. This confirms different interaction cross-sections with the dipole-moment of two-dimensional materials. (b) Polar plot of the PL signal’s integrated intensity at different pump polarizations.}
\end{figure}

\section{Conclusion}
We have demonstrated a new elastomeric hybrid photonic platform where light-matter interaction with WS$_{2}$-hBN, a 2D heterostructure, is facilitated by miniaturization of the waveguide and mode-center integration of the material. Simulations and experiments confirm that the system exhibits significantly enhanced interaction between the encapsulated material and waveguide compared to a previous platform. Most notable is the newly gained ability to detect guide-coupled photoluminescence. The dry-transfer technique used for integration is applicable to other two-dimensional material systems. \par
While semiconductor platforms such as silicon-nitride waveguides can offer higher coupling efficiencies of \raisebox{-0.9ex}{\~{}}7 \%\cite{peyskens_integration_2019} due to a higher index-contrast (\emph{n$_{Si_3N_4}$ }=2.055 , \emph{n$_{SiO_2}$} = 1.461 at 532 nm)\cite{luke_broadband_2015,malitson_interspecimen_1965} and smaller mode area (of \raisebox{-0.9ex}{\~{}}0.14 – 0.2	$\mu$m$^{2}$ ), the presented polymer platform offers unique features such as strain-tunability, low fabrication costs, a short development time, and a simple fabrication procedure. Mild curing conditions also enable versatile approaches to material integration. Besides fundamental studies of TMDs, and other 2D materials, we believe that this system will be of interest in the context of microfluidics, chemical-tuning, flexible photonics, optoelectronic devices, and encapsulation of other nano- and micro-scale materials into polymeric waveguides for novel hybrid systems. While a higher refractive-index contrast of the materials is desirable for further miniaturization of the platform e.g. to build on-chip interferometers and multi-component circuits, the newly demonstrated ability to couple photoluminescence to the waveguide facilitates applications where 2D materials may also be used as integrated (non-)linear and (non-classical) light sources in the near-room-temperature regime. 

\section{Methods}
\subsection{Heterostructure Fabrication and Transfer}
The WS$_{2}$-hBN heterostructure was assembled using a pick-up technique. WS$_{2}$ monolayers were first grown onto Si substrates via a liquid-mediated chemical vapor deposition method. hBN flakes were tape-exfoliated from bulk hBN crystals supplied by National Institute for Materials Science (NIMS), Japan, onto polydimethylsiloxane (PDMS) substrates (Sylgard184, Dow). These hBN flakes were subsequently used to pick-up the CVD-grown WS$_{2}$ flakes. During the pick-up, the WS$_{2}$ samples were heated to 130 $^{\circ}$C to facilitate the pick-up process. Lastly, the WS$_{2}$-hBN heterostack was released onto the waveguide using a dry-transfer technique.

\subsection{Waveguide Fabrication}
Waveguides are fabricated from two different PDMS formulations (Sylgard184 and Gelest OE50) on two silicon wafers. To create the waveguide ridge with mode-center placement, a silicon wafer is spin-coated with a 1 $\mu$m thick layer of positive photoresist (AZ1512). The waveguide pattern is laser-written into this layer. Gelest OE50 is spin-coated onto this mold to a thickness of \raisebox{-0.9ex}{\~{}} 1 $\mu$m. The wafer is cured at 55 °C for 4 hrs and the TMD-hBN heterostructure is transferred via soft lithography, making up the first half of the guiding layer.
A second, unpatterned photoresist-coated substrate is prepared again by spin-coating and curing a 1 $\mu$m thick layer of GelestOE50 on the AZ-coated wafer. A 2 mm thick layer of Sylgard184 (the lower-refractive index polymer) is cured on top of this second substrate at 70 $^{\circ}$C for 1 hr. This substrate now contains the lower part of the guiding layer (Gelest OE50) and its support matrix (Sylgard184). After treatment with oxygen plasma for 30 seconds, the two halves are brought in contact, forming a permanent bond and completing the structure. For off-center placement, we increase the thickness of the high-index polymer-layer to 2 $\mu$m, transfer WS$_{2}$-hBN of the same dimensions as for the mode-center device onto it and plasma-bond the layer directly to the low-index polymer. 

\begin{acknowledgement}
A.L acknowledges support by the National Research Foundation, Prime Minister’s Office, Singapore. 
G.E. acknowledges support from the Ministry of Education (MOE), Singapore, under AcRF Tier 3 (MOE2018-T3-1-005) and the Singapore National Research Foundation for funding the research under the medium-sized center program.
The authors thank Prof. Harald Giessen for the helpful correspondence and for facilitating material characterization and thank Isa Ahmadalidokht for fruitful discussions.

\end{acknowledgement}

\begin{suppinfo}
\subsection{Refractive Index Measurements}
Refractive indices of the polymers were measured using a commercially available critical-angle refractometer (Schmidt-Haensch, ATR-L).
\begin{figure}
 \centering
    \includegraphics[width=17cm]{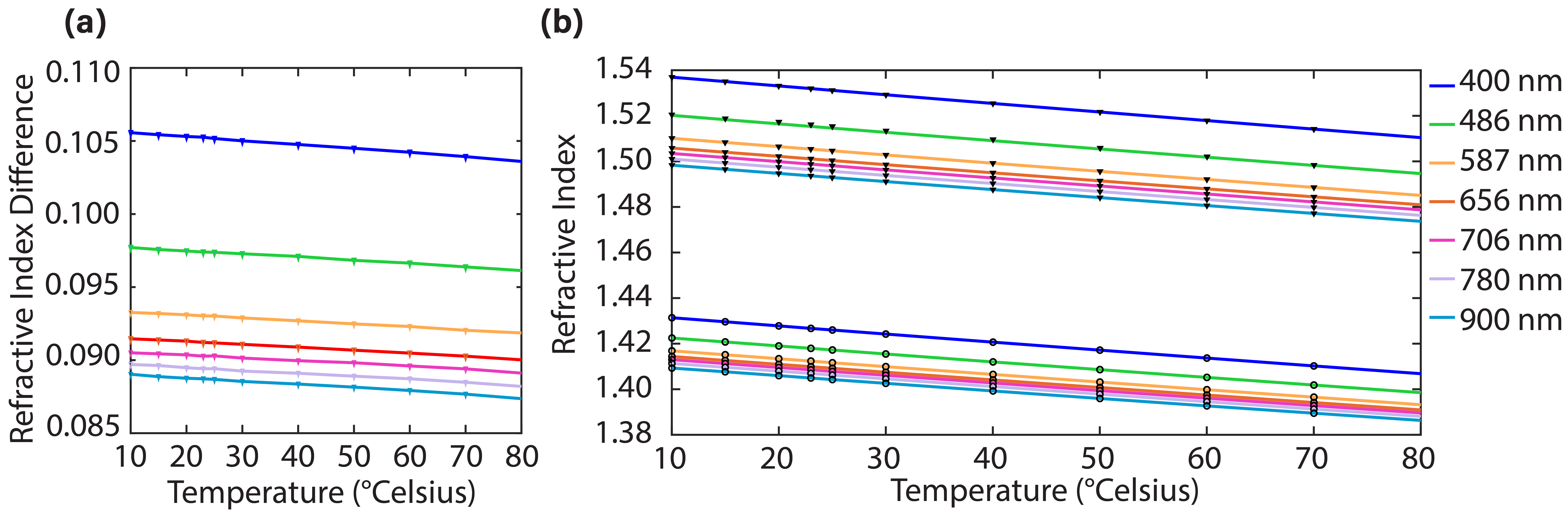}
  \caption{(a) Refractive index difference between high- and low-index polymer (b) Refractive index measurements for the used polymers. Range 1.44-1.38: Sylgard 184; Range 1.54 - 1.47: Gelest OE50.}
\end{figure}

\end{suppinfo}
\par
\bibliography{Bibliography}

\providecommand{\latin}[1]{#1}
\makeatletter
\providecommand{\doi}
  {\begingroup\let\do\@makeother\dospecials
  \catcode`\{=1 \catcode`\}=2 \doi@aux}
\providecommand{\doi@aux}[1]{\endgroup\texttt{#1}}
\makeatother
\providecommand*\mcitethebibliography{\thebibliography}
\csname @ifundefined\endcsname{endmcitethebibliography}
  {\let\endmcitethebibliography\endthebibliography}{}
\begin{mcitethebibliography}{40}
\providecommand*\natexlab[1]{#1}
\providecommand*\mciteSetBstSublistMode[1]{}
\providecommand*\mciteSetBstMaxWidthForm[2]{}
\providecommand*\mciteBstWouldAddEndPuncttrue
  {\def\EndOfBibitem{\unskip.}}
\providecommand*\mciteBstWouldAddEndPunctfalse
  {\let\EndOfBibitem\relax}
\providecommand*\mciteSetBstMidEndSepPunct[3]{}
\providecommand*\mciteSetBstSublistLabelBeginEnd[3]{}
\providecommand*\EndOfBibitem{}
\mciteSetBstSublistMode{f}
\mciteSetBstMaxWidthForm{subitem}{(\alph{mcitesubitemcount})}
\mciteSetBstSublistLabelBeginEnd
  {\mcitemaxwidthsubitemform\space}
  {\relax}
  {\relax}

\bibitem[Malard \latin{et~al.}(2013)Malard, Alencar, Barboza, Mak, and
  de~Paula]{malard_observation_2013}
Malard,~L.~M.; Alencar,~T.~V.; Barboza,~A. P.~M.; Mak,~K.~F.; de~Paula,~A.~M.
  Observation of Intense Second Harmonic Generation from {MoS}$_2$ Atomic
  Crystals. \emph{Physical Review B} \textbf{2013}, \emph{87}, 201401\relax
\mciteBstWouldAddEndPuncttrue
\mciteSetBstMidEndSepPunct{\mcitedefaultmidpunct}
{\mcitedefaultendpunct}{\mcitedefaultseppunct}\relax
\EndOfBibitem
\bibitem[Yin \latin{et~al.}(2014)Yin, Ye, Chenet, Ye, O’Brien, Hone, and
  Zhang]{yin_edge_2014}
Yin,~X.; Ye,~Z.; Chenet,~D.~A.; Ye,~Y.; O’Brien,~K.; Hone,~J.~C.; Zhang,~X.
  Edge {Nonlinear} {Optics} on a {MoS$_2$} {Atomic} {Monolayer}. \emph{Science}
  \textbf{2014}, \emph{344}, 488--490\relax
\mciteBstWouldAddEndPuncttrue
\mciteSetBstMidEndSepPunct{\mcitedefaultmidpunct}
{\mcitedefaultendpunct}{\mcitedefaultseppunct}\relax
\EndOfBibitem
\bibitem[Mak and Shan(2016)Mak, and Shan]{mak_photonics_2016}
Mak,~K.~F.; Shan,~J. Photonics and Optoelectronics of {2D} Semiconductor
  Transition Metal Dichalcogenides. \emph{Nature Photonics} \textbf{2016},
  \emph{10}, 216--226\relax
\mciteBstWouldAddEndPuncttrue
\mciteSetBstMidEndSepPunct{\mcitedefaultmidpunct}
{\mcitedefaultendpunct}{\mcitedefaultseppunct}\relax
\EndOfBibitem
\bibitem[Sun \latin{et~al.}(2016)Sun, Martinez, and Wang]{sun_optical_2016}
Sun,~Z.; Martinez,~A.; Wang,~F. Optical Modulators with {2D} Layered Materials.
  \emph{Nature Photonics} \textbf{2016}, \emph{10}, 227--238\relax
\mciteBstWouldAddEndPuncttrue
\mciteSetBstMidEndSepPunct{\mcitedefaultmidpunct}
{\mcitedefaultendpunct}{\mcitedefaultseppunct}\relax
\EndOfBibitem
\bibitem[Srivastava \latin{et~al.}(2015)Srivastava, Sidler, Allain, Lembke,
  Kis, and Imamoğlu]{srivastava_optically_2015}
Srivastava,~A.; Sidler,~M.; Allain,~A.~V.; Lembke,~D.~S.; Kis,~A.;
  Imamoğlu,~A. Optically Active Quantum Dots in Monolayer {WSe}$_2$.
  \emph{Nature Nanotechnology} \textbf{2015}, \emph{10}, 491--496\relax
\mciteBstWouldAddEndPuncttrue
\mciteSetBstMidEndSepPunct{\mcitedefaultmidpunct}
{\mcitedefaultendpunct}{\mcitedefaultseppunct}\relax
\EndOfBibitem
\bibitem[Koperski \latin{et~al.}(2015)Koperski, Nogajewski, Arora, Cherkez,
  Mallet, Veuillen, Marcus, Kossacki, and Potemski]{koperski_single_2015}
Koperski,~M.; Nogajewski,~K.; Arora,~A.; Cherkez,~V.; Mallet,~P.;
  Veuillen,~J.-Y.; Marcus,~J.; Kossacki,~P.; Potemski,~M. Single Photon
  Emitters in Exfoliated {WSe}$_2$ Structures. \emph{Nature Nanotechnology}
  \textbf{2015}, \emph{10}, 503--506\relax
\mciteBstWouldAddEndPuncttrue
\mciteSetBstMidEndSepPunct{\mcitedefaultmidpunct}
{\mcitedefaultendpunct}{\mcitedefaultseppunct}\relax
\EndOfBibitem
\bibitem[Tran \latin{et~al.}(2017)Tran, Choi, Scott, Xu, Zheng, Seniutinas,
  Bendavid, Fuhrer, Toth, and Aharonovich]{tran_room-temperature_2017}
Tran,~T.~T.; Choi,~S.; Scott,~J.~A.; Xu,~Z.-Q.; Zheng,~C.; Seniutinas,~G.;
  Bendavid,~A.; Fuhrer,~M.~S.; Toth,~M.; Aharonovich,~I. Room-{Temperature}
  {Single}-{Photon} {Emission} from {Oxidized} {Tungsten} {Disulfide}
  {Multilayers}. \emph{Advanced Optical Materials} \textbf{2017}, \emph{5},
  1600939\relax
\mciteBstWouldAddEndPuncttrue
\mciteSetBstMidEndSepPunct{\mcitedefaultmidpunct}
{\mcitedefaultendpunct}{\mcitedefaultseppunct}\relax
\EndOfBibitem
\bibitem[Chu \latin{et~al.}(2015)Chu, Ilatikhameneh, Klimeck, Rahman, and
  Chen]{chu_electrically_2015}
Chu,~T.; Ilatikhameneh,~H.; Klimeck,~G.; Rahman,~R.; Chen,~Z. Electrically
  {Tunable} {Bandgaps} in {Bilayer} {MoS$_2$}. \emph{Nano Letters}
  \textbf{2015}, \emph{15}, 8000--8007\relax
\mciteBstWouldAddEndPuncttrue
\mciteSetBstMidEndSepPunct{\mcitedefaultmidpunct}
{\mcitedefaultendpunct}{\mcitedefaultseppunct}\relax
\EndOfBibitem
\bibitem[Wang \latin{et~al.}(2012)Wang, Kalantar-Zadeh, Kis, Coleman, and
  Strano]{wang_electronics_2012}
Wang,~Q.~H.; Kalantar-Zadeh,~K.; Kis,~A.; Coleman,~J.~N.; Strano,~M.~S.
  Electronics and Optoelectronics of Two-Dimensional Transition Metal
  Dichalcogenides. \emph{Nature Nanotechnology} \textbf{2012}, \emph{7},
  699--712\relax
\mciteBstWouldAddEndPuncttrue
\mciteSetBstMidEndSepPunct{\mcitedefaultmidpunct}
{\mcitedefaultendpunct}{\mcitedefaultseppunct}\relax
\EndOfBibitem
\bibitem[Rivera \latin{et~al.}(2018)Rivera, Yu, Seyler, Wilson, Yao, and
  Xu]{rivera_interlayer_2018}
Rivera,~P.; Yu,~H.; Seyler,~K.~L.; Wilson,~N.~P.; Yao,~W.; Xu,~X. Interlayer
  Valley Excitons in Heterobilayers of Transition Metal Dichalcogenides.
  \emph{Nature Nanotechnology} \textbf{2018}, \emph{13}, 1004--1015\relax
\mciteBstWouldAddEndPuncttrue
\mciteSetBstMidEndSepPunct{\mcitedefaultmidpunct}
{\mcitedefaultendpunct}{\mcitedefaultseppunct}\relax
\EndOfBibitem
\bibitem[Mouri \latin{et~al.}(2013)Mouri, Miyauchi, and
  Matsuda]{mouri_tunable_2013}
Mouri,~S.; Miyauchi,~Y.; Matsuda,~K. Tunable {Photoluminescence} of {Monolayer}
  {MoS$_2$} via {Chemical} {Doping}. \emph{Nano Letters} \textbf{2013},
  \emph{13}, 5944--5948\relax
\mciteBstWouldAddEndPuncttrue
\mciteSetBstMidEndSepPunct{\mcitedefaultmidpunct}
{\mcitedefaultendpunct}{\mcitedefaultseppunct}\relax
\EndOfBibitem
\bibitem[Castellanos-Gomez \latin{et~al.}(2013)Castellanos-Gomez, Roldán,
  Cappelluti, Buscema, Guinea, van~der Zant, and
  Steele]{castellanos-gomez_local_2013}
Castellanos-Gomez,~A.; Roldán,~R.; Cappelluti,~E.; Buscema,~M.; Guinea,~F.;
  van~der Zant,~H. S.~J.; Steele,~G.~A. Local {Strain} {Engineering} in
  {Atomically} {Thin} {MoS$_2$}. \emph{Nano Letters} \textbf{2013}, \emph{13},
  5361--5366\relax
\mciteBstWouldAddEndPuncttrue
\mciteSetBstMidEndSepPunct{\mcitedefaultmidpunct}
{\mcitedefaultendpunct}{\mcitedefaultseppunct}\relax
\EndOfBibitem
\bibitem[Elshaari \latin{et~al.}(2020)Elshaari, Pernice, Srinivasan, Benson,
  and Zwiller]{elshaari_hybrid_2020}
Elshaari,~A.~W.; Pernice,~W.; Srinivasan,~K.; Benson,~O.; Zwiller,~V. Hybrid
  Integrated Quantum Photonic Circuits. \emph{Nature Photonics} \textbf{2020},
  \emph{14}, 285--298\relax
\mciteBstWouldAddEndPuncttrue
\mciteSetBstMidEndSepPunct{\mcitedefaultmidpunct}
{\mcitedefaultendpunct}{\mcitedefaultseppunct}\relax
\EndOfBibitem
\bibitem[Sibson \latin{et~al.}(2017)Sibson, Erven, Godfrey, Miki, Yamashita,
  Fujiwara, Sasaki, Terai, Tanner, Natarajan, Hadfield, O’Brien, and
  Thompson]{sibson_chip-based_2017}
Sibson,~P.; Erven,~C.; Godfrey,~M.; Miki,~S.; Yamashita,~T.; Fujiwara,~M.;
  Sasaki,~M.; Terai,~H.; Tanner,~M.~G.; Natarajan,~C.~M.; Hadfield,~R.~H.;
  O’Brien,~J.~L.; Thompson,~M.~G. Chip-Based Quantum Key Distribution.
  \emph{Nature Communications} \textbf{2017}, \emph{8}, 13984\relax
\mciteBstWouldAddEndPuncttrue
\mciteSetBstMidEndSepPunct{\mcitedefaultmidpunct}
{\mcitedefaultendpunct}{\mcitedefaultseppunct}\relax
\EndOfBibitem
\bibitem[Flamini \latin{et~al.}(2018)Flamini, Spagnolo, and
  Sciarrino]{flamini_photonic_2018}
Flamini,~F.; Spagnolo,~N.; Sciarrino,~F. Photonic Quantum Information
  Processing: A Review. \emph{Reports on Progress in Physics} \textbf{2018},
  \emph{82}, 016001\relax
\mciteBstWouldAddEndPuncttrue
\mciteSetBstMidEndSepPunct{\mcitedefaultmidpunct}
{\mcitedefaultendpunct}{\mcitedefaultseppunct}\relax
\EndOfBibitem
\bibitem[Chamanzar \latin{et~al.}(2013)Chamanzar, Xia, Yegnanarayanan, and
  Adibi]{chamanzar_hybrid_2013}
Chamanzar,~M.; Xia,~Z.; Yegnanarayanan,~S.; Adibi,~A. Hybrid Integrated
  Plasmonic-Photonic Waveguides for on-Chip Localized Surface Plasmon Resonance
  ({LSPR}) Sensing and Spectroscopy. \emph{Optics Express} \textbf{2013},
  \emph{21}, 32086--32098\relax
\mciteBstWouldAddEndPuncttrue
\mciteSetBstMidEndSepPunct{\mcitedefaultmidpunct}
{\mcitedefaultendpunct}{\mcitedefaultseppunct}\relax
\EndOfBibitem
\bibitem[Kohler \latin{et~al.}(2021)Kohler, Schindler, Hahn, Milvich, Hofmann,
  Länge, Freude, and Koos]{kohler_biophotonic_2021}
Kohler,~D.; Schindler,~G.; Hahn,~L.; Milvich,~J.; Hofmann,~A.; Länge,~K.;
  Freude,~W.; Koos,~C. Biophotonic Sensors with Integrated
  {Si}$_3${N}$_4$-Organic Hybrid ({SiNOH}) Lasers for Point-of-Care
  Diagnostics. \emph{Light: Science \& Applications} \textbf{2021}, \emph{10},
  64\relax
\mciteBstWouldAddEndPuncttrue
\mciteSetBstMidEndSepPunct{\mcitedefaultmidpunct}
{\mcitedefaultendpunct}{\mcitedefaultseppunct}\relax
\EndOfBibitem
\bibitem[Porcel \latin{et~al.}(2019)Porcel, Hinojosa, Jans, Stassen, Goyvaerts,
  Geuzebroek, Geiselmann, Dominguez, and Artundo]{porcel_invited_2019}
Porcel,~M. A.~G.; Hinojosa,~A.; Jans,~H.; Stassen,~A.; Goyvaerts,~J.;
  Geuzebroek,~D.; Geiselmann,~M.; Dominguez,~C.; Artundo,~I. {Silicon} Nitride
  Photonic Integration for Visible Light Applications. \emph{Optics \& Laser
  Technology} \textbf{2019}, \emph{112}, 299--306\relax
\mciteBstWouldAddEndPuncttrue
\mciteSetBstMidEndSepPunct{\mcitedefaultmidpunct}
{\mcitedefaultendpunct}{\mcitedefaultseppunct}\relax
\EndOfBibitem
\bibitem[Tonndorf \latin{et~al.}(2017)Tonndorf, Del Pozo-Zamudio, Gruhler,
  Kern, Schmidt, Dmitriev, Bakhtinov, Tartakovskii, Pernice, Michaelis~de
  Vasconcellos, and Bratschitsch]{tonndorf_-chip_2017}
Tonndorf,~P.; Del Pozo-Zamudio,~O.; Gruhler,~N.; Kern,~J.; Schmidt,~R.;
  Dmitriev,~A.~I.; Bakhtinov,~A.~P.; Tartakovskii,~A.~I.; Pernice,~W.;
  Michaelis~de Vasconcellos,~S.; Bratschitsch,~R. On-{Chip} {Waveguide}
  {Coupling} of a {Layered} {Semiconductor} {Single}-{Photon} {Source}.
  \emph{Nano Letters} \textbf{2017}, \emph{17}, 5446--5451\relax
\mciteBstWouldAddEndPuncttrue
\mciteSetBstMidEndSepPunct{\mcitedefaultmidpunct}
{\mcitedefaultendpunct}{\mcitedefaultseppunct}\relax
\EndOfBibitem
\bibitem[Joshi and Kaushik(2020)Joshi, and Kaushik]{joshi_transition_2020}
Joshi,~S.; Kaushik,~B.~K. Transition Metal Dichalcogenides Integrated Waveguide
  Modulator and Attenuator in Silicon Nitride Platform. \emph{Nanotechnology}
  \textbf{2020}, \emph{31}, 435202\relax
\mciteBstWouldAddEndPuncttrue
\mciteSetBstMidEndSepPunct{\mcitedefaultmidpunct}
{\mcitedefaultendpunct}{\mcitedefaultseppunct}\relax
\EndOfBibitem
\bibitem[Ngo \latin{et~al.}(2020)Ngo, George, Schock, Tuniz, Najafidehaghani,
  Gan, Geib, Bucher, Knopf, Saravi, Neumann, Lühder, Schartner,
  Warren‐Smith, Ebendorff‐Heidepriem, Pertsch, Schmidt, Turchanin, and
  Eilenberger]{ngo_scalable_2020}
Ngo,~G.~Q. \latin{et~al.}  Scalable {Functionalization} of {Optical} {Fibers}
  {using} {Atomically} {Thin} {Semiconductors}. \emph{Advanced Materials}
  \textbf{2020}, \emph{32}, 2003826\relax
\mciteBstWouldAddEndPuncttrue
\mciteSetBstMidEndSepPunct{\mcitedefaultmidpunct}
{\mcitedefaultendpunct}{\mcitedefaultseppunct}\relax
\EndOfBibitem
\bibitem[Zuo \latin{et~al.}(2020)Zuo, Yu, Liu, Cheng, Qiao, Liang, Zhou, Wang,
  Wu, Zhao, Gao, Wu, Sun, Liu, Bai, and Liu]{zuo_optical_2020}
Zuo,~Y. \latin{et~al.}  Optical Fibres with Embedded Two-Dimensional Materials
  for Ultrahigh Nonlinearity. \emph{Nature Nanotechnology} \textbf{2020},
  \emph{15}, 987--991\relax
\mciteBstWouldAddEndPuncttrue
\mciteSetBstMidEndSepPunct{\mcitedefaultmidpunct}
{\mcitedefaultendpunct}{\mcitedefaultseppunct}\relax
\EndOfBibitem
\bibitem[Vogl \latin{et~al.}(2017)Vogl, Lu, and Lam]{vogl_room_2017}
Vogl,~T.; Lu,~Y.; Lam,~P.~K. Room Temperature Single Photon Source using
  Fiber-Integrated Hexagonal Boron Nitride. \emph{Journal of Physics D: Applied
  Physics} \textbf{2017}, \emph{50}, 295101\relax
\mciteBstWouldAddEndPuncttrue
\mciteSetBstMidEndSepPunct{\mcitedefaultmidpunct}
{\mcitedefaultendpunct}{\mcitedefaultseppunct}\relax
\EndOfBibitem
\bibitem[Peyskens \latin{et~al.}(2019)Peyskens, Chakraborty, Muneeb,
  Van~Thourhout, and Englund]{peyskens_integration_2019}
Peyskens,~F.; Chakraborty,~C.; Muneeb,~M.; Van~Thourhout,~D.; Englund,~D.
  Integration of Single Photon Emitters in {2D} Layered Materials with a
  Silicon Nitride Photonic Chip. \emph{Nature Communications} \textbf{2019},
  \emph{10}, 4435\relax
\mciteBstWouldAddEndPuncttrue
\mciteSetBstMidEndSepPunct{\mcitedefaultmidpunct}
{\mcitedefaultendpunct}{\mcitedefaultseppunct}\relax
\EndOfBibitem
\bibitem[He \latin{et~al.}(2021)He, Paradisanos, Liu, Cadore, Liu, Churaev,
  Wang, Raja, Javerzac-Galy, Roelli, Fazio, Rosa, Tongay, Soavi, Ferrari, and
  Kippenberg]{he_low-loss_2021}
He,~J. \latin{et~al.}  Low-{Loss} {Integrated} {Nanophotonic} {Circuits} with
  {Layered} {Semiconductor} {Materials}. \emph{Nano Letters} \textbf{2021},
  \emph{21}, 2709--2718\relax
\mciteBstWouldAddEndPuncttrue
\mciteSetBstMidEndSepPunct{\mcitedefaultmidpunct}
{\mcitedefaultendpunct}{\mcitedefaultseppunct}\relax
\EndOfBibitem
\bibitem[Iff \latin{et~al.}(2019)Iff, Tedeschi, Martín-Sánchez,
  Moczała-Dusanowska, Tongay, Yumigeta, Taboada-Gutiérrez, Savaresi,
  Rastelli, Alonso-González, Höfling, Trotta, and
  Schneider]{iff_strain-tunable_2019}
Iff,~O.; Tedeschi,~D.; Martín-Sánchez,~J.; Moczała-Dusanowska,~M.;
  Tongay,~S.; Yumigeta,~K.; Taboada-Gutiérrez,~J.; Savaresi,~M.; Rastelli,~A.;
  Alonso-González,~P.; Höfling,~S.; Trotta,~R.; Schneider,~C.
  Strain-{Tunable} {Single} {Photon} {Sources} in {WSe$_2$} {Monolayers}.
  \emph{Nano Letters} \textbf{2019}, \emph{19}, 6931--6936\relax
\mciteBstWouldAddEndPuncttrue
\mciteSetBstMidEndSepPunct{\mcitedefaultmidpunct}
{\mcitedefaultendpunct}{\mcitedefaultseppunct}\relax
\EndOfBibitem
\bibitem[Errando-Herranz \latin{et~al.}(2021)Errando-Herranz, Schöll, Picard,
  Laini, Gyger, Elshaari, Branny, Wennberg, Barbat, Renaud, Sartison,
  Brotons-Gisbert, Bonato, Gerardot, Zwiller, and
  Jöns]{errando-herranz_resonance_2021}
Errando-Herranz,~C. \latin{et~al.}  Resonance {Fluorescence} from
  {Waveguide}-{Coupled}, {Strain}-{Localized}, {Two}-{Dimensional} {Quantum}
  {Emitters}. \emph{ACS Photonics} \textbf{2021}, \emph{8}, 1069--1076\relax
\mciteBstWouldAddEndPuncttrue
\mciteSetBstMidEndSepPunct{\mcitedefaultmidpunct}
{\mcitedefaultendpunct}{\mcitedefaultseppunct}\relax
\EndOfBibitem
\bibitem[Kim \latin{et~al.}(2020)Kim, Aghaeimeibodi, Carolan, Englund, and
  Waks]{kim_hybrid_2020}
Kim,~J.-H.; Aghaeimeibodi,~S.; Carolan,~J.; Englund,~D.; Waks,~E. Hybrid
  Integration Methods for on-Chip Quantum Photonics. \emph{Optica}
  \textbf{2020}, \emph{7}, 291--308\relax
\mciteBstWouldAddEndPuncttrue
\mciteSetBstMidEndSepPunct{\mcitedefaultmidpunct}
{\mcitedefaultendpunct}{\mcitedefaultseppunct}\relax
\EndOfBibitem
\bibitem[Auksztol \latin{et~al.}(2019)Auksztol, Vella, Verzhbitskiy, Ng, Ho,
  Grieve, Viana-Gomes, Eda, and Ling]{auksztol_elastomeric_2019}
Auksztol,~F.; Vella,~D.; Verzhbitskiy,~I.; Ng,~K.~F.; Ho,~Y.~W.; Grieve,~J.~A.;
  Viana-Gomes,~J.; Eda,~G.; Ling,~A. Elastomeric {Waveguide} on-{Chip}
  {Coupling} of an {Encapsulated} {MoS$_2$} {Monolayer}. \emph{ACS Photonics}
  \textbf{2019}, \emph{6}, 595--599\relax
\mciteBstWouldAddEndPuncttrue
\mciteSetBstMidEndSepPunct{\mcitedefaultmidpunct}
{\mcitedefaultendpunct}{\mcitedefaultseppunct}\relax
\EndOfBibitem
\bibitem[Pérez-Calixto \latin{et~al.}(2017)Pérez-Calixto,
  Zamarrón-Hernández, Cruz-Ramírez, Hautefeuille, Hérnandez-Cordero,
  Velázquez, and Grether]{perez-calixto_fabrication_2017}
Pérez-Calixto,~D.; Zamarrón-Hernández,~D.; Cruz-Ramírez,~A.;
  Hautefeuille,~M.; Hérnandez-Cordero,~J.; Velázquez,~V.; Grether,~M.
  Fabrication of Large All-{PDMS} Micropatterned Waveguides for Lab on Chip
  Integration using a Rapid Prototyping Technique. \emph{Optical Materials
  Express} \textbf{2017}, \emph{7}, 1343--1350\relax
\mciteBstWouldAddEndPuncttrue
\mciteSetBstMidEndSepPunct{\mcitedefaultmidpunct}
{\mcitedefaultendpunct}{\mcitedefaultseppunct}\relax
\EndOfBibitem
\bibitem[Shabahang \latin{et~al.}(2021)Shabahang, Clouser, Shabahang, and
  Yun]{shabahang_single-mode_2021}
Shabahang,~S.; Clouser,~F.; Shabahang,~F.; Yun,~S.-H. Single-{Mode},
  700\%-{Stretchable}, {Elastic} {Optical} {Fibers} {Made} of {Thermoplastic}
  {Elastomers}. \emph{Advanced Optical Materials} \textbf{2021}, \emph{9},
  2100270\relax
\mciteBstWouldAddEndPuncttrue
\mciteSetBstMidEndSepPunct{\mcitedefaultmidpunct}
{\mcitedefaultendpunct}{\mcitedefaultseppunct}\relax
\EndOfBibitem
\bibitem[Kee \latin{et~al.}(2009)Kee, Poenar, Neuzil, and
  Yobas]{kee_design_2009}
Kee,~J.~S.; Poenar,~D.~P.; Neuzil,~P.; Yobas,~L. Design and Fabrication of
  {Poly}(dimethylsiloxane) Single-Mode Rib Waveguide. \emph{Optics Express}
  \textbf{2009}, \emph{17}, 11739--11746\relax
\mciteBstWouldAddEndPuncttrue
\mciteSetBstMidEndSepPunct{\mcitedefaultmidpunct}
{\mcitedefaultendpunct}{\mcitedefaultseppunct}\relax
\EndOfBibitem
\bibitem[Missinne \latin{et~al.}(2014)Missinne, Kalathimekkad, Hoe, Bosman,
  Vanfleteren, and Steenberge]{missinne_stretchable_2014}
Missinne,~J.; Kalathimekkad,~S.; Hoe,~B.~V.; Bosman,~E.; Vanfleteren,~J.;
  Steenberge,~G.~V. Stretchable Optical Waveguides. \emph{Optics Express}
  \textbf{2014}, \emph{22}, 4168--4179\relax
\mciteBstWouldAddEndPuncttrue
\mciteSetBstMidEndSepPunct{\mcitedefaultmidpunct}
{\mcitedefaultendpunct}{\mcitedefaultseppunct}\relax
\EndOfBibitem
\bibitem[Grieve \latin{et~al.}(2017)Grieve, Ng, Rodrigues, Viana-Gomes, and
  Ling]{grieve_mechanically_2017}
Grieve,~J.~A.; Ng,~K.~F.; Rodrigues,~M. J. L.~F.; Viana-Gomes,~J.; Ling,~A.
  Mechanically Tunable Integrated Beamsplitters on a Flexible Polymer Platform.
  \emph{Applied Physics Letters} \textbf{2017}, \emph{111}, 211106\relax
\mciteBstWouldAddEndPuncttrue
\mciteSetBstMidEndSepPunct{\mcitedefaultmidpunct}
{\mcitedefaultendpunct}{\mcitedefaultseppunct}\relax
\EndOfBibitem
\bibitem[Schuller \latin{et~al.}(2013)Schuller, Karaveli, Schiros, He, Yang,
  Kymissis, Shan, and Zia]{schuller_orientation_2013}
Schuller,~J.~A.; Karaveli,~S.; Schiros,~T.; He,~K.; Yang,~S.; Kymissis,~I.;
  Shan,~J.; Zia,~R. Orientation of Luminescent Excitons in Layered
  Nanomaterials. \emph{Nature Nanotechnology} \textbf{2013}, \emph{8},
  271--276\relax
\mciteBstWouldAddEndPuncttrue
\mciteSetBstMidEndSepPunct{\mcitedefaultmidpunct}
{\mcitedefaultendpunct}{\mcitedefaultseppunct}\relax
\EndOfBibitem
\bibitem[Gissibl \latin{et~al.}(2017)Gissibl, Wagner, Sykora, Schmid, and
  Giessen]{gissibl_refractive_2017}
Gissibl,~T.; Wagner,~S.; Sykora,~J.; Schmid,~M.; Giessen,~H. Refractive index
  measurements of photo-resists for three-dimensional direct laser writing.
  \emph{Optical Materials Express} \textbf{2017}, \emph{7}, 2293--2298\relax
\mciteBstWouldAddEndPuncttrue
\mciteSetBstMidEndSepPunct{\mcitedefaultmidpunct}
{\mcitedefaultendpunct}{\mcitedefaultseppunct}\relax
\EndOfBibitem
\bibitem[Schmid \latin{et~al.}(2019)Schmid, Ludescher, and
  Giessen]{schmid_optical_2019}
Schmid,~M.; Ludescher,~D.; Giessen,~H. Optical properties of photoresists for
  femtosecond {3D} printing: refractive index, extinction, luminescence-dose
  dependence, aging, heat treatment and comparison between 1-photon and
  2-photon exposure. \emph{Optical Materials Express} \textbf{2019}, \emph{9},
  4564--4577\relax
\mciteBstWouldAddEndPuncttrue
\mciteSetBstMidEndSepPunct{\mcitedefaultmidpunct}
{\mcitedefaultendpunct}{\mcitedefaultseppunct}\relax
\EndOfBibitem
\bibitem[Luke \latin{et~al.}(2015)Luke, Okawachi, Lamont, Gaeta, and
  Lipson]{luke_broadband_2015}
Luke,~K.; Okawachi,~Y.; Lamont,~M. R.~E.; Gaeta,~A.~L.; Lipson,~M. Broadband
  Mid-Infrared Frequency Comb Generation in a
  {Si}$_{\textrm{3}}${N}$_{\textrm{4}}$ Microresonator. \emph{Optics Letters}
  \textbf{2015}, \emph{40}, 4823--4826\relax
\mciteBstWouldAddEndPuncttrue
\mciteSetBstMidEndSepPunct{\mcitedefaultmidpunct}
{\mcitedefaultendpunct}{\mcitedefaultseppunct}\relax
\EndOfBibitem
\bibitem[Malitson(1965)]{malitson_interspecimen_1965}
Malitson,~I.~H. Interspecimen {Comparison} of the {Refractive} {Index} of
  {Fused} {Silica}. \emph{JOSA} \textbf{1965}, \emph{55}, 1205--1209\relax
\mciteBstWouldAddEndPuncttrue
\mciteSetBstMidEndSepPunct{\mcitedefaultmidpunct}
{\mcitedefaultendpunct}{\mcitedefaultseppunct}\relax
\EndOfBibitem
\end{mcitethebibliography}

\end{document}